\documentclass[10pt,letterpaper]{article}
\usepackage[top=0.85in,left=2.75in,footskip=0.75in]{geometry}

\usepackage{amsmath,amssymb}

\usepackage{changepage}

\usepackage[utf8x]{inputenc}

\usepackage{textcomp,marvosym}

\usepackage{cite}

\usepackage{nameref,hyperref}

\usepackage[right]{lineno}

\usepackage{microtype}
\DisableLigatures[f]{encoding = *, family = * }

\usepackage[table]{xcolor}

\usepackage{array}

\newcolumntype{+}{!{\vrule width 2pt}}

\newlength\savedwidth

\raggedright
\usepackage[aboveskip=1pt,labelfont=bf,labelsep=period,justification=raggedright,singlelinecheck=off]{caption}

\makeatletter
\renewcommand{\@biblabel}[1]{\quad#1.}
\makeatother

\usepackage{lastpage,fancyhdr,graphicx}
\usepackage{epstopdf}
\fancyheadoffset[L]{2.25in}
\fancyfootoffset[L]{2.25in}
\begin{document}
\vspace*{0.2in}

\begin{flushleft}
{\Large
\textbf\newline{Building Trust: Lessons from the Technion-Rambam Machine Learning in Healthcare Datathon Event} 
}
\newline
\\
Jonathan A. Sobel\textsuperscript{1},
Ronit Almog\textsuperscript{2},
Leo Anthony Celi\textsuperscript{3},
Michal Gaziel-Yablowitz\textsuperscript{4},
Danny Eytan\textsuperscript{2},
Joachim A. Behar\textsuperscript{1}*
\\
\bigskip
\textbf{1} Faculty of Biomedical Engineering, Technion, Israel Institute of Technology, Haifa, Israel
\\
\textbf{2} Rambam Health Care Campus, Haifa, Israel
\\
\textbf{3} Institute for Medical Engineering and Science, Massachusetts Institute of Technology, Department of Medicine, Beth Israel Deaconess Medical Center; Department of Biostatistics, Harvard T.H. Chan School of Public Health, Boston, MA, 02115, USA
\\
\textbf{4} TIMNA- Israel's Ministry of Health  Big Data Platform, The Ministry of Health
\\
\bigskip

* jbehar@technion.ac.il

\end{flushleft}
\section*{Abstract}
A joint conference and datathon event organized by the Technion, Rambam Health Care Campus and the MIT Critical Data group took place on March 7-9th 2022 and provided a unique opportunity to work with real-world clinical datasets and open clinical questions that may impact medical practices. Clinicians from Rambam posed challenges that they face in their practice. Participants mostly originated  from six different faculties and were divided into teams with mixed skillsets. Each team was assigned a research project as well as a senior clinical mentor and a data scientist mentor. At the end of the second day of the competition, three teams were shortlisted and presented their analysis at the final plenary session of a conference held the following day in front of a panel of senior technologists, machine learning (ML) researchers and health professionals. The datathon demonstrated the potential in crowd sourcing of bright minds of students from different faculties and backgrounds to formulate innovative solutions that can impact the clinical practice. It facilitated definition of a roadmap for data preparation and safe data sharing by the epidemiology and the IT units of the hospital. It helped understand how opening these data to research can bear significant impacts on the management  of the hospital patients and beyond. Most importantly, this event served as a vector for establishment of a dialogue to bridge between a the Technion, a technology institute, and Rambam, its natural partnering hospital, in the field of medical data science. By demonstrating to stakeholders from both institutions the potential in such data sharing and team work, this datathon will likely catalyze future collaborations.  We plan to strengthen this bridge through the newly established joint academic Technion-Rambam Center for Artificial Intelligence in Healthcare (CAIH).

\section*{Author summary}
A datathon is a time-constrained competition involving data science applied to a specific problem. In the past decade, datathons have been shown to be a valuable bridge between fields and expertise. Biomedical data analysis represents a challenging area requiring collaboration between engineers, biologists and physicians to gain a better understanding of patient physiology and of guide decision processes for diagnosis, prognosis and therapeutic interventions to improve care practice. Here, we reflect on the outcomes of an event that we organized in Israel at the end of March 2022 between the MIT Critical Data group, Rambam Health Care Campus (Rambam) and the Technion Israel Institute of Technology (Technion) in Haifa. Participants were asked to complete a survey about their skills and interests, which enabled us to identify current needs in machine learning training for medical problem applications. This work describes opportunities and limitations in medical data science in the Israeli  context.

\nolinenumbers

\section*{Introduction}
\subsection*{Context of the event}
This joint event organized by the Technion, Rambam and the MIT Critical Data group provided a unique opportunity to understand the challenges faced by leading researchers and clinicians working in the field of medical data science. The Technion is one of the world’s leading science and technology research institutes and Rambam is the largest hospital in the north of Israel. The event took place in the city of Haifa, Israel on both the Technion and Rambam campuses. It was organized as the inaugural event of a new joint Technion-Rambam Center for Artificial Intelligence in Healthcare (CAIH), which aims to serve as an academic center for medical AI committed to advanced medical and clinical research, with significant and actionable benefit to patient care. This first joint academic-hospital AI center in Israel, plans to develop advanced AI systems for patient evaluation driven by complex and rapid analysis of all the relevant medical information that has accumulated in big medical databases over the years \cite{Sapci2020}. The Center’s opening event entitled “Technion-Rambam Hack: Machine Learning in Healthcare,” was held on March 7-9th 2022 and was attended by about 250 people. The first two days consisted of a collaborative information-based competition \cite{lyndon2018hacking,serpa2018first,pathanasethpong2017tackling,li2017promoting} that  focused on solving real-world clinical problems through interdisciplinary teams and access to real data \cite{aboab2016datathon}. The datathon was followed by a one-day conference with lectures delivered by researchers from the Technion, Rambam, MIT, the Israeli Ministry of Health (MOH), Clalit Health Services, GE Healthcare, and Roche. Scientists, healthcare practitioners and policy makers from across the globe shared their knowledge on the fascinating topic of medical AI.

\subsection*{The complex structure of Israel’s healthcare system}
Israel’s healthcare system is based on various organizations with diverse types of ownership and complex relations between them \cite{naamati2020strategic}. Each Israeli resident is entitled to healthcare coverage provided by one of four non-profits health maintenance organizations (HMOs)- Clalit, Maccabi, Leumit and Meuhedet, which compete with each other \cite{naamati2020strategic}. The HMOs deliver quality community services close to the patient’s home and use various payment methods to purchase services supplied by hospitals \cite{clarfield2017health}. There are general hospitals across Israel with diverse types of ownership: Clalit owns 14 hospitals, MOH and non-profit organizations own 17 hospitals and Maccabi owns two hospitals. At the same time, the MOH serves as the owner of 11 general hospitals, the regulator that oversees all public hospital expenses and the legislator. Until 1994, Clalit and Leumit provided care to the majority of Israel’s residents who were unionized in two separate labour organizations, while Maccabi and Meuhedet acted as private health providers and had the privilege to define their own admissions requirements. As a result, 5\% of the population was not covered by health insurance, and Clalit, the largest HMO in the country, accumulated huge debts, which jeopardized its ability to provide care to its members. The legislation of the 1994 National Health Insurance Law changed Israel’s healthcare system dramatically. It granted minimum essential healthcare coverage to all Israeli residents and the right to choose a healthcare provider. The reforms in public healthcare services had far-reaching financial ramifications. New agreements and fee updates increased the competition among HMOs and decreased the HMOs’ use of hospital services \cite{naamati2020strategic}. As a result, hospital directors adopted a business-oriented attitude in order to generate income beyond their usual financial support from the state. The need to reduce operational costs and increase the value of care led hospitals and HMOs to computerize their electronic health records (EHR) and invest in digital infrastructure for their storage and sharing \cite{jones2019promoting}.

\subsection*{Stakeholders hold different levels of data} 

Israel is a world leader in use of health data to reduce costs and complexity, while increasing patient satisfaction and outcomes. For 25 years already, Israel’s hospitals and HMOs use different EHR platforms, with each type of healthcare organization holding access to different levels of health data. All hospitals hold full EHRs, that include clinical measurements, lab test results, images, text (reports made by nurses or physicians) and medications prescribed during the patient’s visit. Community-based clinics hold longitudinal health data that include medical history, lab results of routine check-ups, routine medication, images taken in the clinics, and administrative data and costs. Each public hospital owned by the MOH or non-profit organization collects and stores data separately. Community-based clinics affiliated with Clalit or Maccabi collect and share in-network data with their affiliated hospitals. Meuhedet or Leumit share in-network community-based data without the ability to link it with full hospital data. Massive volumes of data about patients, conditions and treatments have accumulated over the years. Very recently, the Israeli health authorities decided to invest \$30M in 19 hospital programs that promoted anonymized data sharing between healthcare organizations. To support some degree of data sharing between all stakeholders, the MOH implemented the ”Ofek ” system, which is a secured health information exchange network that enables hospital teams to view patient medical history, routine medications, test results, and imaging and pathology data in HMO records during their hospitalization \cite{gerber2014promoting}. In addition, all hospitals are required to deliver detailed information to the MOH’s hospitalization database. The collected information includes, socio-demographic details, medical diagnostic data and administrative data from each patient visit in every Israeli hospital. Moreover, the MOH owns more than 20 different population-level clinical datasets, which includes perinatal and child developmental data up to the age of six, births and death causes databases, the COVID-19 database and other national registries. In addition, the MOH owns the TIMNA platform (Research Infrastructure for Big Data), a national framework that supports data sharing across all types of health organization, and enables access to the MOH’s national databases, through a neutral, virtual and secured research environment \cite{fabiana2017trends,peleg2021collaboration,jaffe2020role,world2020country}.

\vspace{5mm}

Overall, we organised this event to serve as a vector for establishment of a dialogue to bridge between a the Technion, a technology institute, and Rambam, its natural partnering hospital, in the field of medical data science. By demonstrating to stakeholders from both institutions and beyond (academia, HMO, companies, governmental organisations and non-governmental organisations), the potential in such data sharing and team work, this datathon will likely catalyze future collaborations in Israel. 

\section*{Methods}

\section*{How to organize a datathon?}
To organize a successful event, several important points should be well thought through before the event. The checklist provided below should help any organizer in this process. One of the first decisions is related to the place of the future event, should it be virtual or in person. Next, a list of potential partners should be assembled to help fund the event and attract participants. Leaders in the field who will present some of their work are an asset to the event. Moreover, some of these partners can play the role of mentors during the competition. Another critical point for a datathon involves finding questions of clinical importance that can be addressed using previously collected data. There are several options here as some events may be more flexible in the sense that participants/projects leader can come with their data and questions or a more directed approach with defined datasets and questions. In all cases, the data should be properly anonymized and the ethical statements from the institutional review board (IRB) should be provided before the event in the case of medical data. When medical data are used a secure infrastructure such as AWS or Azure can be set up to avoid any leakage of medical data. Mentors that can follow each team during the whole event are necessary to ensure the success of each project. The choice of the targeted audience is one of the main points. The goal of a datathon is to demonstrate the effectiveness of a multidisciplinary approach where each team member can provide different expertise (medical, technical, social, legal and business). There are two options there: 1) a more directed approach where teams are built by the organizers from a pool of participants or 2) the self-organization approach where participants choose their project before or at the beginning of the event. Last, the reward for the winning teams is important to acknowledge teams that have performed well as well as encourage them continuing the research afterward. Participants are generally striving for learning new skills and networking but the reward will increase their motivation to succeed.

\vspace{5mm}

\textbf{Datathon Check list}
\begin{itemize}
\item Venue: Physical/Virtual/Hybrid, dates, location.
\item  Logistics: catering, strong WIFI, rooms and amphitheater.
\item  Partners: Industrial, NGO, clinicians and academic stakeholders , who may fund some part of the event (awards/venue/infrastructure) as well as deliver relevant talks during the event.
\item  Projects: Research project call for datasets with ethical consents (IRB) and specified aims/questions. 
\item  IT support and secured computational infrastructure (for sensitive clinical data to be shared with participants).
\item  Mentors: Clinical and data science. Try to select senior mentors.
\item  Participants: Who is your targeted audience? (students/medical professionals/data scientists)
\item  Communication/PR: Information to participants and advertisement of the event (Flyers/website/social media)
\item  Awards: Money for the winning teams or other gifts

\end{itemize}

\subsection*{The Datathon days}

The Datathon is an information-based competition \cite{aboab2016datathon}. It was held at the Technion Faculty of Biomedical Engineering. Teams were organized to be cross-disciplinary, and included data scientists and students from healthcare disciplines (biomedical engineering, medicine, biology, physics, electrical	engineering, and computer science). The	planning of the datathon and the conference began approximately six months before the event	(Figure 1). Several tasks were planned in each phase of the event organization. After initial brainstorming between the scientific committee, which included Technion principal scientists, Rambam clinicians and MIT scientists, a fundraising campaign was launched as list of potential speakers for the conference day was drawn up and invitations were extended. Communication around the event was initiated in November 2021 via social media platforms (Twitter, LinkedIn and Facebook). Students interested in the datathon were asked to apply to the event and were asked to complete a survey  about their skills, their interests and their level of education (Bsc., Msc., Ph.D., alumni) and specialty (engineering or bio/med). In parallel, we contacted	clinicians from Rambam and asked them to propose projects consisting of a medical question and to provide a relevant dataset to research the question. Four challenges proposed by clinicians who had collected large datasets in recent years and who presented challenging scientific questions which could be tackled by ML were selected. The projects were 1) Prediction of newborn birthweight by maternal parameters and previous newborn siblings birthweights  \cite{mccowan2005customised}, 2) ML-based  predictive model for bloodstream infections during hematopoietic stem cell transplantation \cite{gupta2020systematic}, 3) Prediction of recurrent hospitalization in heart failure patients  \cite{ouwerkerk2014factors} and 4) Risk  factor and severity prediction in hospitalized Covid-19 patients  \cite{benaim2021comparing,Sobel2021.09.26.21264135}. Each challenge came from a different hospital department and was presented with a database of approximately 2 Gb data. The data were anonymized and structured by the Rambam epidemiology and IT teams \cite{o2004health,kruse2017security}. The ethical approval by the institutional review board to work on these datasets was checked and updated, if necessary. The four challenges relied on EHR data (i.e. structured data) \cite{johnson2016mimic,hyland2020early}. This was preferred over the usage of unstructured data (time series or images) because of the format of the Datathon which is two days long, and because of constraint on the budget for cloud computing.

Clinicians who proposed the	project were, by default, the medical mentors and two competing teams composed of 5-7 participants were assigned to each project. A senior data scientist was recruited for each team either from Technion labs, or Rambam epidemiology department or industry partners. The role of the mentors was to advise and monitor their team’s activities throughout the entire event. Participants were selected based on their interests and competency (studies and skills). Our goal was to have mixed teams in terms of data analysis capacity and field knowledge to work on each challenge. Next, the computational infrastructure was built in collaboration with the Rambam IT team	using Amazon Web Services (AWS) \cite{fusaro2011biomedical} and tests were run to simulate the event. Our requirement was to work on a secure cloud that provides data access without data leak or data export to a participant’s computer. We started with a minimal configuration of an 8-core machine with 16Gb RAM for each team. During the event, teams were allowed access to more computational resources, on request. Each team	was asked to test the cloud two weeks before the event and was allowed to set up their	infrastructure with programs and libraries (e.g., R, Python). In a second testing phase, the internet access from the cloud was closed and teams were allowed to explore the structure of the data using a small sample of patients. Finally, one day before the event, the full databases were loaded to the cloud. Each team had a separate virtual machine with personal, secured access for each team member. During the two days of the event, the teams were split in several rooms at the Technion Faculty	of Biomedical Engineering. After the first day’s lunch, a networking event in the form of a scientific speed dating was held with mentors, team members and organizers, to promote future collaborations. Each team was asked to present its work at the end of the second day. Thereafter, using an external jury comprised of a principal investigator from the Technion, clinicians, Rambam epidemiology and IT department, and industrial partners, the three best  teams were selected for the competition final, which took place on the conference day.	

\subsection*{The Conference day}

The guest talks were intended to introduce the field of clinical data science to a wide audience, including those with both technical and non-technical backgrounds, and  to provide a perspective on its future impact on the field of medicine. For this purpose, the conference day was structured into three main plenary sessions. Session 1 entitled “Current trends in ML in healthcare” addressed recent research in medical ML in the fields of omics, medical image analysis, causal inference and physiological time series analysis \cite{biton2021atrial,levy2021digital,chocron2020remote,NEURIPS2021_ffa4eb0e}. The presentations began with a general introduction and presented examples of research questions that can be tackled by ML for the purpose of diagnosis, risk prediction and treatment as well as basic science. Session 2 was entitled “Data stakeholders” and aimed to present the medical data landscape in Israel, and to clearly present the different actors in Israel i.e. Hospitals, HMOs and MOH and the data availability options, limitations, types of partnerships and costs. The speakers emphasized which data are available at each level (Hospitals, HMOs and MOH) as well as the bridges that exist between the different actors. The speakers also discussed how their entity collaborates with academia and industry. Session 3 entitled “Deployment of ML in medical practice” discussed how ML algorithms translate to medical practice. In other words, how ML models trained on retrospective data can be translated to clinical practice, including algorithm integration, training and perception of medical staff, importance of distribution shifts and model generalization \cite{zhang2022interactive}.

\begin{figure}[!h]
\includegraphics[scale=.7]{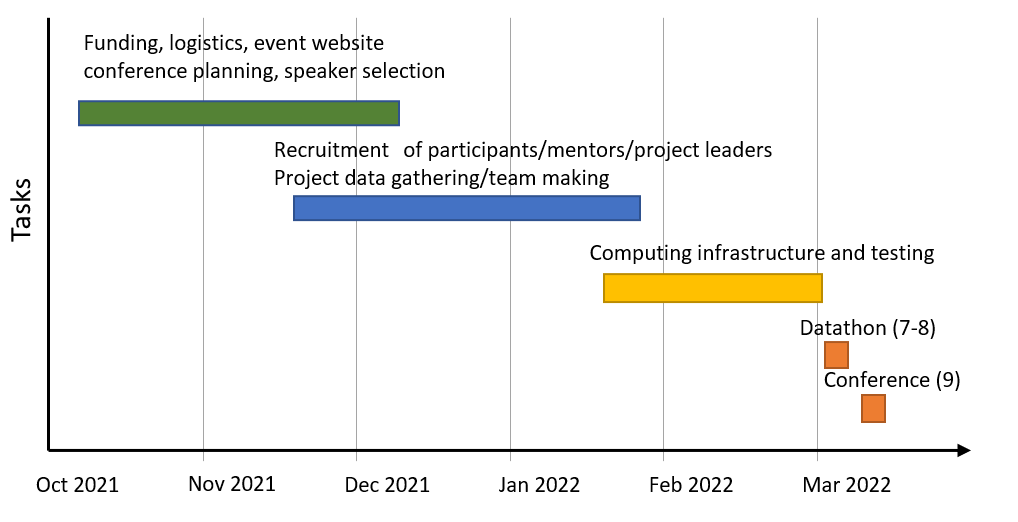}
\caption{{\bf Datathon and conference preparation and execution timeline.}
}
\label{fig1}
\end{figure}

\begin{figure}[!h]
\includegraphics[scale=.9]{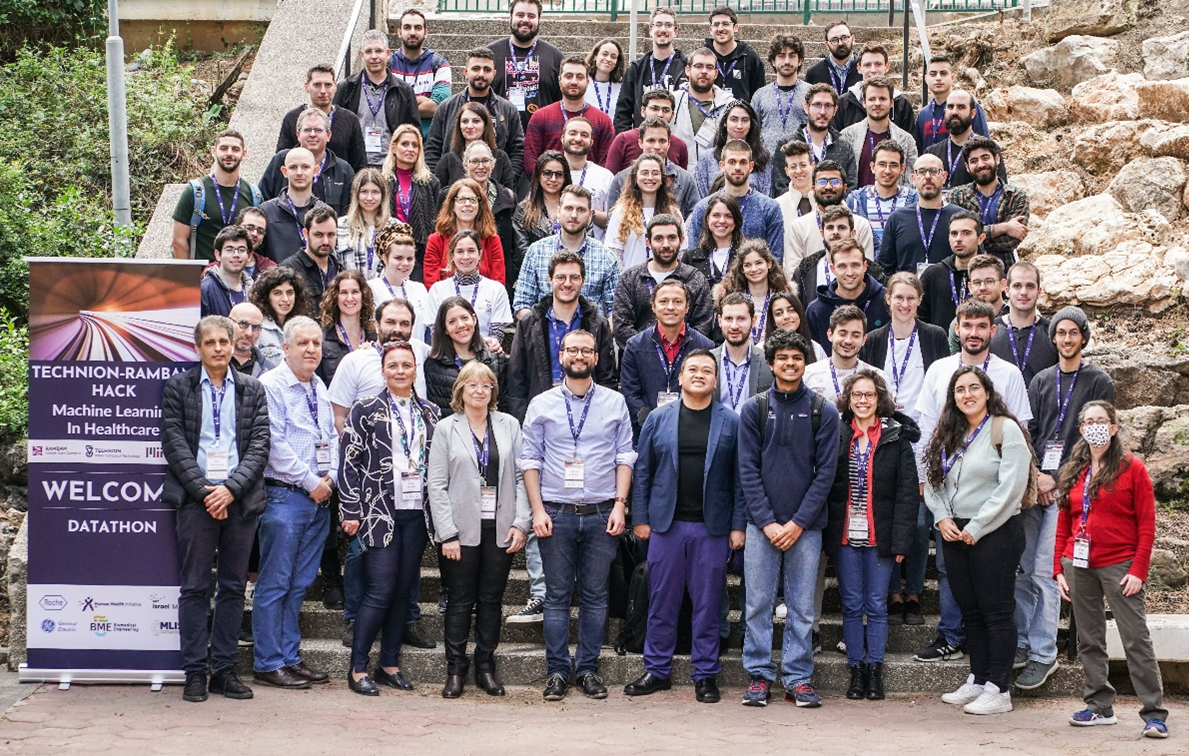}
\caption{{\bf Datathon's participants and mentors.}
A total of 50 students and alumni participated in the event and were advised by 20 mentors.
Pictures copyright Nitzan Zohar (Technion)
}
\label{fig2}
\end{figure}

\section*{Results}

The datathon was attended by 50 students and alumni from various Technion faculties. Participants developed different strategies for the analysis of medical data, to solve important challenges in cardiology, fetal monitoring, intensive care, and stem cell transplantation. The students were accompanied by 20 mentors from the Technion, Rambam and the industry. Among the 80 applicants, most were engineers (75\%) and BSc students (50\%). Consequently, in order to organize teams with diverse backgrounds and education levels, we favoured participants with MSc or PhD and with a bio/medical background. Thus, of the selected 50 participants, 20\% where PhD, 40\% were MSc and 40\% where BSc (Figure 3A). We asked participants to report their field of study, their favourite programming language and their medical field of interest and to rate their competency in various technical skills (Figure 3B-E). Approximately 33\% of the participants reported that they had a bio/medical oriented background (Figure 3B), and the majority of engineers came from the biomedical engineering faculty of the Technion. Interestingly, the next largest subgroup was computer scientists. The preferred programming language was Python, as this language is well-suited for ML, free and well-studied across many Technion faculties (Figure 3C). Many participants reported on good knowledge of Matlab, as many courses at the Technion rely on this licensed computational tool. Only a minority had previous knowledge of R and other languages used in data-science research. Participants were interested in several medical fields (Figure 3D), with a slight preference for radiology, oncology and cardiology. Participants reported having a good background of data analysis and data-visualization and ML in general (Figure 3E). They reported slightly lower proficiency in biostatistics, modelling and image analysis. Finally, participants were less confident in the analysis of EHR, database management, time series and text analysis.

\begin{figure}[!h]
\includegraphics[scale=1.3]{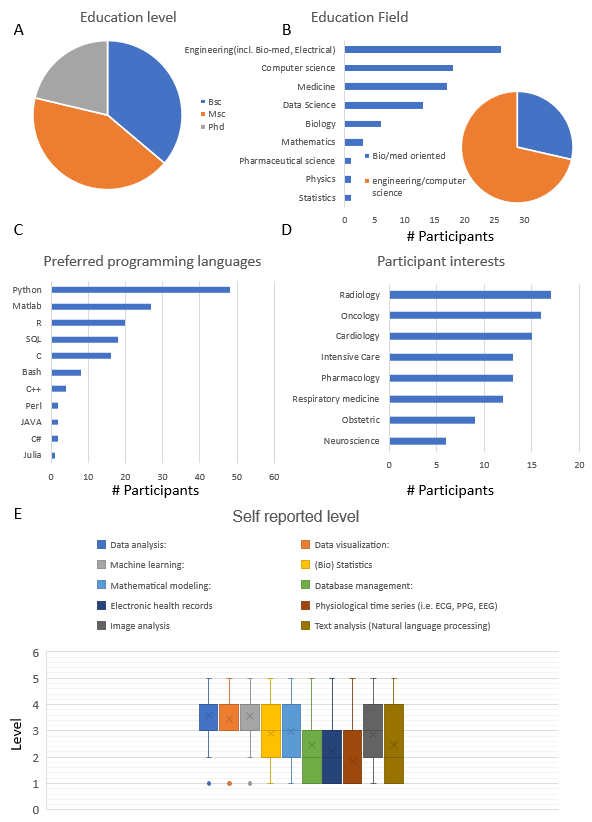}
\caption{{\bf Survey results.}  A) Level of education, B) specialty, C) programming skills, D) medical field of interest and E) data analysis and ML skills. Note that participants could provide multiple answers for their orientation, skills and interests.
}
\label{fig3}
\end{figure}

\begin{figure}[!h]
\includegraphics[scale=.56]{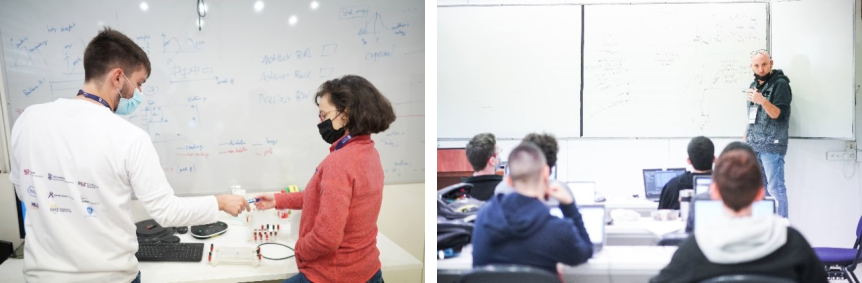}
\caption{{\bf Participants and mentors interaction.} Collaboration between students during the datathon event (left image) and a medical mentor explaining the meaning of some of the variables to one of the datathon team. Pictures copyright Nitzan Zohar (Technion).
}
\label{fig4}
\end{figure}

\section*{Discussion} 
As arose from participant surveys, this type of event attracts the young generation of engineers, who are used to collaborative projects, open source and open data initiatives. On the other hand, there were fewer bio/med student applicants. This may be because they were less confident about their contribution due to their lack of coding/data management skills. However, we observed that successful teams were those comprised of clinicians (physicians, nurses), data managers, hospital IT \& infrastructure officer, and engineers (developers, medical device, data acquisition). These teams achieved a better understanding of the pathologies and challenges faced in the clinic. Furthermore, this datathon event underscored the potential in exploiting the bright minds of university students from various faculties and backgrounds to design solutions that can impact clinical practice. It facilitated defining a road for data preparation and safe data sharing by the epidemiology and the IT units of the hospital. It helped understand how opening medical data to research can significantly impact patient management and beyond. Most importantly, the event seemed to be the vector necessary for establishing a dialogue between the Technion and its natural partnering hospital (Rambam) in the field of medical data science. Leveraging such a meeting between a technological university and its nearby hospital institution can serve as a model of trust building between partners. We will strengthen this bridge through the newly established CAIH.

\subsection*{Barriers to development and adoption of ML tools in medicine}
There are multiple barriers to development and, moreover, adoption of ML tools in medicine that span the full cycle, from inception to deployment. Some relate to data availability as detailed here and others to technological difficulties, culture- and local ecosystem-related resistance, and lack of an enabling infrastructure in most Institutions, many of which we plan to address as part of the mission of CAIH. Other barriers stem from a chasm between the clinical domain experts and their ML counterparts; development of accurate, robust, actionable, and fair ML tools requires close collaboration between clinical domain experts, data scientists and engineers. For such collaboration to be truly fruitful, these specialists must be able to interact and exchange ideas and thoughts using a common language and basic understanding of key concepts from both worlds. Typically, most clinicians and data scientists lack this reciprocally needed knowledge. Specifically, most current-day medical training programs in Israel and abroad do not include courses dedicated to advanced data analytics, such as introduction to ML.

Moreover, in most medical schools, even if students wish to expand their knowledge on ML and its applications to healthcare, there is no option to do so. Of note, in this Regard, the Technion is unique in that it has established several undergraduate programs for a dual bachelor’s degree in medicine and either computer science or biomedical engineering, which can be completed before clinical training begins. These programs require two additional years of study. It is our view though, that considering the expected near-future integration of multiple artificial-intelligence-based technologies into everyday practice of most clinicians, a basic understanding of concepts related to how such algorithms work, their advantages and failure modes will be advantageous and even necessary. Such basic knowledge is required to appropriately operate these tools, similar to the way a clinician must understand how an electrocardiogram signal is generated to be able to accurately diagnose common and rare diseases or recognize data acquisition problems. Thus, we believe that medical schools should be cognizant of the changing landscape and adjust their curricula accordingly, to include some basic training in ML. The flip-side of this challenge lies in the need to equip engineering students who plan to specialized in ML for healthcare with basic medical knowledge and familiarity with relevant medical context.  Many leading institutions, including the Technion, established specific courses dedicated to introducing engineering students to this field and its unique jargon, challenges, and progress. Additional initiatives in this direction include datathons, such as the one described here, and promotion of joint undergraduate and graduate projects dedicated to ML in healthcare and spanning engineering and medical faculties. In that sense, institutions such as the Technion, which is a technology
institute yet also has a large medical faculty, bring an advantage.

\subsection*{Accessing medical data in Israel: a critical appraisal}

Despite the fact that Israel likely has some of the best medical datasets in the world, due to the structuring of its healthcare system and the very early digitization of personal medical data, access to medical data remains challenging due to ethical concerns and regulation (data protection, anonymisation, GDPR, HIPAA) and their potential business value (intellectual property). This has important consequences on the Israeli academy seeking to access these data. In addition, data fragmentation poses a major challenge. There are three main data stakeholders in Israel: the HMOs, the hospitals and the MOH. Each with their own agenda. Challenges arise when when seeking to use data across multiple stakeholders, e.g., imaging data from a hospital, with clinical endpoints from the MOH or longitudinal data which is only be available at the HMO level. There is currently no ”one-stop shop” address for collection of all available data. The final challenge is the infrastructure. It is our belief that the state should be independent and sovereign when it comes to medical data of citizens. In this respect, we believe that Israel should have its own, i.e., physically located in Israel and solely owned by the government, high-performance computing (HPC) facilities hosting medical data for research and development. This is with the exception of temporary hosting of medical data on classical clouds or international platforms in the context of federated learning. New initiatives by the MOH, such as TIMNA, or the new program to promote anonymized data sharing between healthcare organizations, may tackle some of these challenges. Yet, new regulations will be necessary for the MOH to allow medical data sharing for research purposes in Israeli universities. Without this legal enforcement, the business focus of health institutions may remain preponderant and limit the academic research performed in Israel.

\section*{Conclusion}

This datathon was very attractive and built trust and establish a pipeline for medical data preparation and sharing between medical and technology institutions. It provided a critical appraisal of education needs for health professionals and engineers seeking to pursue ML fields. Finally, this work represents a first field test for medical data access in Israel. Obstacles are numerous but Israel has the potential to exist as a main actor in the field of medical ML because of its strong basis in engineering and medicine and its history of medical data digitization.

\section*{Ethical statement}
Statistics were provided as part of the application process to the datathon and were anonymized. Each participants signed an non-disclosure agreement stating that any collected data for the challenges and during the event belong to the Rambam HCC. Participants to the event were told that pictures would be taken during the event and could be used for marketing purposes of the event.


\section*{Acknowledgments}
We acknowledge the financial support of the Technion-Rambam Center for Artificial Intelligence in Medicine, the Technion Machine Learning and Intelligent Systems  (MLIS) Center, the Technion Human Health Initiative (THHI) and the MISTI MIT - Israel Zuckerman STEM Fund. This research was partially supported by The Milner Foundation, founded by Yuri Milner and his wife Julia. We are grateful to the Placide Nicod Foundation for their financial support (J.S.). We are grateful to the medical  and technical staff from the IT and epidemiology departments at Rambam HCC. We are thankful to the Technion administrative staff for supporting the organization and communication of the Datathon event.

\nolinenumbers

%
%
%
\bibliography{biblio}

\end{document}